\documentstyle[12pt,epsf]{article}
\begin{document}
\title{Gauge Symmetry from Integral Viewpoint}
\author{{Ying Chen${}^{2}$\thanks{Email address: cheny@hptc5.ihep.ac.cn},
         Bing He${}^{2}$\thanks{Email address: heb@alpha02.ihep.ac.cn}, He
Lin${}^{2}$, Ji-min Wu${}^{1,2}$\thanks{Email address:
wujm@alpha02.ihep.ac.cn}}\\
{\small ${}^{1}$ CCAST (World Laboratory), P.O.Box 8730, Beijing
100080, P.R.China}\\
   {\small ${}^{2}$ Institute of High Energy Physics, Academia Sinica, 
Beijing
100039, P.R.China}}
\maketitle
\begin{center}
\begin{minipage}{140mm}
\vskip 0.8in
\begin{center}{\bf Abstract}\end{center}
{The purpose of this paper is to investigate the gauge symmetry of classical 
field theories in integral formalism. A gauge invariant theory is 
defined in terms of the invariance of the physical observables under the 
coordinate transformations in principal bundle space. Through the detailed 
study on the properties of non-Abelian parallel transporter under gauge 
transformations, we show that it is not generally a two-point spinor, i.e. 
an operator to be affected only by the gauge group elements at the two end 
points of the parallel transport path, except for the pure gauge situation, 
and therefore the local gauge symmetry for non-Abelian models is found to be 
broken in non-perturbative domain. However, an Abelian gauge theory is proved 
to be strictly invariant under local gauge transformation, as it is illustrated 
by the invariance of the interference pattern of electrons in Aharonov-Bohm 
effect. The related issues of the phenomenon
are discussed.
}
\end{minipage}
\end{center}
\baselineskip 0.3in
\newpage
\par
\section{Introduction}
It has been accepted without doubt that the present-day physical theories 
of the
fundamental interactions in nature largely follow from the principle
of gauge symmetry, which postulates that any physically acceptable
field model should be invariant under gauge transformations.
For a concrete gauge field model it is constructed from the principle as 
follows:
1) The requirement for the invariance of the field action $S$ under the action
of a finitely dimensional Lie group leads to the existence of conserved
currents (Noether theorem); 2) The generalization of the invariance of the
action under the global gauge transformations to the local ones introduces in
the coupling of the currents through a gauge field. In fact it is just the
second step, i.e. the replacement of $\partial_{\mu}$ in action by $D_{\mu}$, 
that enables us to have complete theories to describe the matter fields and the
interactions coupling matters. The quantized gauge field model
(standard model)
has been successful to a large extent in explaining and predicting the
experimental results in elementary particle physics---for an overview, see e.g. Ref.[1]. 

The invariance of field action, or more exactly
the invariance of Lagrangian density under gauge transformations,
constitutes the basis for the
differential formalism of a gauge field theory. According to it, two field
models
$\langle \it L, A_{\mu}, \psi \rangle$ and $\langle \it L', A'_{\mu}, \psi'
\rangle$ are equivalent if they are related to each other by the form invariance of Lagrangian
density $\it L$, and therefore by the invariance of local field equations,
under gauge transformations including the global (GT1) and
local (GT2a for matter field and GT2b for gauge field) ones. Here we have used 
the notations in Ref.[2].
It is assumed as a matter of course that other formalisms of gauge field 
theory, e.g. the integral
formalism [3-4], should be the same as the differential one with 
respect to the inherent gauge symmetry. To meet the requirement, the gauge transformation
of a parallel transporter, the counterpart of gauge field $A_{\mu}(x)$ in
the integral formalism, should be a {\bf two-point
spinor}\footnote[1]{It is
termed in
analogy to {\bf two-point tensor}~in [5-6] for the parallel 
transporter of a tensor field in spacetime manifold.}, i.e. that to be 
affected only by
the gauge transformation GT2a at the two end points which the parallel
transporter connects. Contrary to this assumption, however,
we prove that the two-point spinor property for a parallel 
transporter for a non-Abelian gauge field is conditional if the field
is non-trivial and, therefore, the gauge
freedom for a non-Abelian theory is much more restricted than what we see from
differential point of view on gauge theories. 

In this paper, we first clarify in the framework of fibre bundle some
commonly used terminologies
such as gauge transformation and gauge invariance, then give a definition of
gauge invariant field theory in terms of the invariance of physical
observables. After proving 
the equivalence of gauge transformation GT2b and
two-point spinor property of infinitesimal parallel transporter in the form of
one-order expansion, we will study
in detail the property of finite parallel transporter 
in non-Abelian and Abelian gauge theories. Finally, some related issues
will be discussed. 

\section{Gauge transformation and gauge invariance}

In this section we begin with the clarification of some terminologies
regarding a gauge field theory, which are by no means uniform throughout
the literature. We here adopt the definition of gauge theory in Ref.[2]:
\begin{quotation}

    {\bf A gauge field theory} is the one
that is derived from the gauge principle\noindent \footnote[2]
{The above-mentioned
gauge symmetry principle is close to the definition of
{\bf gauge postulate} in Ref.[2], and should be distinguished from the term
{\bf gauge principle} there.}
and represents the geometry of a principal fibre bundle.
The gauge group is given by the structure group of the bundle.
\end{quotation}
Since the geometrical framework of fibre bundle provides a natural
mathematical setting for the representation of physical gauge theories,
we will largely use the language for the clarification of the terminologies
we will use in our study.

A principal fibre bundle is defined as a tuple $\langle E, M, \pi,
G \rangle$ with bundle space $E$, base space $M$, projection map
$\pi$: $E \to M$, and structure group
$G$ which is homeomorphic to fibre space $F$. The base space $M$ we
study in this paper is Minkowskian or Euclidean spacetime for simplicity. 
In fibre bundle language a
matter field is given as a cross section in the associated vector bundle of
principal bundle. It is a smooth map $\psi$: $M \to E$, which satisfies
$\pi \circ \psi(x)=x$ for $\forall x \in M$. Of course there is no
continuous cross section unless the principle bundle is with a trivial topology,
i.e. the product of base and fibre space. To describe gauge field
$A_{\mu}(x)$
(throughout the paper $A_{\mu}$ is the short handed symbol for
$\sum \limits^{dimG}_{a=1}A_{\mu}^aT^a$, where $T^a$s are the generators
of the Lie
algebra of gauge group),
we need to identify $i\lambda A_{\mu}(x)$, where $\lambda$ is the coupling
constant, with
the local representation $B_{\mu}(x)$~[7, p.~160] of connection form
$\omega$,
which is given as [8, p.~255]
\begin {eqnarray}
\omega=-g^{-1}(x)dg(x)+g^{-1}(x)B_{\mu}(x)g(x)dx^{\mu}
\end{eqnarray}
in the local coordinate $\varphi$: $\left( g(x),x \right)$ ( $g \in G$, 
$x \in M$), and which determines a unique decomposition $TE = PE \oplus
HE$
of the tangent space $TE$ of the principal bundle $E$
into a `perpendicular' and a `horizontal' part.

Cross section at each point $x \in U_\alpha$, an arbitrary neighborhood in
base space, can be regarded locally as an element of gauge group. A choice of 
a cross section is equivalent to the choice of a local coordinate, 
and the cross section in a local moving frame $\{e_i\}$ is given as
$\psi(x)=\{\psi^i(x)\}$.
Therefore the local 
gauge transformations GT2a and GT2b can be derived from the transformation of
local coordinates
\begin {eqnarray}
T_{\alpha\beta}=\varphi^{-1}_{\beta}\circ\varphi_{\alpha}:  G \times U_{\alpha
\beta} \to G \times U_{\alpha\beta}
\end{eqnarray}
Under this transformation the cross section
$\psi(x)$ in bundle space
$E_{\alpha}$ transforms as
\begin {eqnarray}
\psi'(x)=g(x)\psi(x)
\end{eqnarray}
with the local moving frame transformation in the
tangent space of the associate vector bundle of
principal bundle: $\{e_i\}\to\{e'_i\}$.
Thus we have got the gauge transformation GT2a. To obtain gauge
transformation
GT2b, we will make use of connection form Eq.(1). In two different
coordinates
$(g_{1}(x),x)$ and $(g_{2}(x),x)$, if a connection form
is expressed as [8, p.~256]
$$\omega_{E_{\alpha}}=-g^{-1}_1(x)dg_1(x)+g^{-1}_1(x)B^{(1)}_{\mu}(x)g_1(x)dx^{\mu}$$
\begin{eqnarray}
~~~~~~~~~~~~=-g^{-1}_2(x)dg_2(x)+g^{-1}_2(x)B^{(2)}_{\mu}(x)g_2(x)dx^{\mu},
\end{eqnarray}
then the relation of the local representations of connection form is
therefore
found to be
\begin {eqnarray}
B^{(2)}_\mu(x)=-g(x)\frac{\partial g^{-1}}{\partial
x^{\mu}}(x)+g(x)B^{(1)}_{\mu}(x)g^{-1}(x),
\end{eqnarray}
where $g=g_1g^{-1}_2$.
Gauge transformation GT2b has been derived in this way if the local representation
$B_{\mu}(x)$ is identified with the gauge field $i\lambda A_{\mu}$.
From Eq. (3) and Eq. (5) we obviously see that the local coordinate
transformation Eq. (2) can be realized by any $C^{1}$ gauge group element 
$g(x)$
(with continuous first derivatives with respect to spacetime coordinates 
$x$), which also preserves the invariance of the field action constructed
by
$\psi(x)$ and $A_{\mu}(x)$. In this sense we call a $C^1$ coordinate
transformation $T_{\alpha\beta}(x)$ a {\bf `general local gauge
transformation'}. 

In Eq. (3) the action of gauge group is performed in the same fibre
over each point in the base space,
so we call it {\bf `perpendicular action'} of gauge group. With connection
form $\omega$ we can also define another type of action of gauge group,
that is, parallel transport of fibres from one point in base space to another.
It involves different points in base space, so it is termed {\bf
`horizontal
action'} of gauge group. In this paper we denote a parallel transporter, the
gauge group element which implements such action, as
$\Phi_{\gamma}(B_{\mu}; x,x_0)$, where the transport path $\gamma$ is
an arbitrary piecewise smooth curve which connects two end points,
say $x_0$ and $x$, and is parametrized by the path variable $t$.
As a point in fibre space, cross section $\psi(x)$ is transported accordingly:
\begin {eqnarray}
\psi(x)=\Phi_{\gamma}(B_{\mu}; x,x_0)\psi(x_0)
\end{eqnarray}
Parallel transport is determined by the `horizontal direction' of the 
bundle space
[8, p.~253]:
\begin {eqnarray}
\omega=-\Phi^{-1}d\Phi+\Phi^{-1}B_{\mu}\Phi dx^\mu=0.
\end{eqnarray}
Over a smooth segment of the path on which there is a definite 
tangent vector field,
$dx^{\mu}(t)/dt$, it is equivalent to the following 
matrix differential equation:
\begin{eqnarray}
\frac{d\Phi}{dt}(t)=B(t)\Phi(t),
\end{eqnarray}
\begin {eqnarray}
\Phi(t_0)=I,
\end{eqnarray}
where $\Phi(t)=\Phi(B(t);t,t_0)$, and $B(t)=B_{\mu}(x(t))
dx^{\mu}(t)/dt$.
There are 
two types of the solution to the equation:
$$\Phi(t)=
Pexp\left(\int^{t}_{t_0}dsB(s)\right)~=I+\int^{t}_{t_0}dsB(s)$$
   \begin{eqnarray}
~+\int^{t}_{t_0}dsB(s)\int^s_{t_0}ds_1B(s_1)+\cdots+
\int^{t}_{t_0}dsB(s)\cdots\int^{s_{n-1}}_{t_0}
   ds_{n}B(s_{n})+\cdots,
\end {eqnarray}   
for the case of non-Abelian gauge group, and 
\begin{eqnarray}
\Phi(t)=exp\left(\int^{t}_{t_0}dsB(s)\right)
\end {eqnarray}
for the case of Abelian gauge group, since the differential equation
in the latter case reduces to an ordinary linear differential equation. A
parallel
transporter along a piecewise
smooth curve or a gauge group element of `horizontal action' is therefore 
given as
the product of these operators defined on the monotonic smooth segments.
Naturally there is the accompanying
differential equation for parallel transport of matter field:
\begin{eqnarray}
\frac {d\psi}{dt}(t)
-i\lambda
A_{\mu}(x(t))\frac{dx^{\mu}}{dt}(t)\psi(t)=D_{\mu}\psi(x(t))\frac{dx^{\mu}}{dt}(t)=0,
\end {eqnarray} 
\begin{eqnarray}
\psi(t_0)=\psi_0.
\end {eqnarray} 
From these results it is
concluded that,
under a parallel transport, the total wave function for a matter field not only
changes its amplitude but also undergoes a `rotation'
in the internal space such as isospin space in $SU(2)$ gauge theory.

Parallel transporter plays an important role in the integral formalism of
gauge
theory. It adequately describes all physics contained
in the gauge theory and, moreover, the summation of it is a physically measurable
quantity because it defines the transition amplitude of a particle moving
along a classical trajectory in the presence of gauge field $A_\mu$. For
overviews of gauge field theories with $\Phi=Pexp\oint_{C}A_{\mu}(x)dx^{\mu}$
as the dynamical variable (loop space formalism), see e.g.
Ref.[9, chap.~7], Ref.[10, chap.~4]

We are now in a position to define gauge invariance, the central concept
in gauge field theory. Following the definition of gauge field theory 
in Ref.[2]
cited at the beginning of the section, we give the definition as follows:
\begin{quotation}
 {\bf A gauge invariant field theory} is a gauge field theory which 
is invariant under the bundle coordinate transformation $T_{\alpha\beta}$,
that is,
all
the observables including the local and non-local ones should be invariant
under its induced transformations GT1 and GT2. The
classification of global and
local invariant theories is according to whether $T_{\alpha\beta}$ is
spacetime
dependent or not.
\end{quotation}
Obviously from this definition we
see that the gauge freedom of a gauge field theory is comprised of the actions
of gauge group that preserve the invariance of the theory.
The distinction of the definition from those elsewhere lies in the invariance
of the non-local operators, such as 
$\bar \psi(x_2)\Phi_{\gamma}(A_{\mu};x_2,x_1)\psi(x_1)$, the operator
for the bound state wave function [10].
Should the operator be invariant under local gauge transformations, the parallel
transporter must be a two-point spinor, i.e. it transforms as follows:
\begin{eqnarray}
\Phi_{\gamma}(A'_{\mu};x_2,x_1)=g(x_2)\Phi_{\gamma}(A_{\mu};x_2,x_1)g^{-1}(x_1),
\end{eqnarray}
under the local gauge transformations.
This two-point spinor property of $\Phi_{\gamma}$ is regarded as
the natural consequence of the general covariance in bundle space (see
e.g.[12]), since
the physics should be independent of the choice of coordinate.
In fibre bundle language we can say that the commutation of
`perpendicular action' and
`horizontal action' of gauge group guarantees a coordinate-free
description
of physics contained in a gauge theory. 
In addition to the gauge fixing terms which are introduced in the
Lagrangian density for the appropriate
physical situations, however, we find that this requirement
actually imposes more restriction on the gauge
freedom of a gauge theory, because we will prove later that it is
conditional in non-Abelian gauge theories.

\section{Joint of differential and integral formalism}

In practice a parallel transporter $\Phi_{\gamma}(x,x_0)$ can be expressed as infinite
product of local operator too. It is obtained through discretization of
Eq. (8):
\begin{eqnarray}
\frac{\Phi(t+\delta)-\Phi(t)}{\delta}=B(t)\Phi(t)
\end{eqnarray}
The solution of the difference equation is
\begin{eqnarray}
\Phi(t+n\delta)=\lim_{n \rightarrow \infty}{\prod\limits_{i=n-1}^0}\left(I+\delta
\it B(i\delta)\right)~\Phi(t_0).
\end{eqnarray}
If there is a definite tangent vector at each point of a segment of the
parallel transport path, we get the alternative form for a 
parallel transporter over the segment:
\begin{eqnarray}
Pexp\int_{t_0}^{t}dsA(s)=\lim_{\Delta t \rightarrow 0}
(I+A(t_{n-1})\Delta t)\cdots (I+
A(t_0)\Delta t),
\end{eqnarray}
where $A(t_i)=i\lambda A_{\mu}(x(t_i))dx^{\mu}(t_i)/dt$, $i=1,2,\cdots,n-1$.

Infinitesimal parallel transporter $I+i\lambda A_{\mu}dx^{\mu}$ bridges over
the connection between differential and integral formalism of gauge theory.
When $\Delta x^{\mu}\rightarrow 0$, one-order approximation 
\begin{eqnarray}
g(x+\Delta x)\approx g(x)+\frac{\partial
g(x)}{\partial x^\mu}\Delta x^{\mu}
\end{eqnarray}
can be regarded to be an exact equality,
and therefore we find that infinitesimal parallel transporter, 
$I+i\lambda A_{\mu}dx^{\mu}$, transforms as
a two-point spinor under gauge transformation GT2b:
$$I+i\lambda A'_{\mu}dx^{\mu}=\left(g(x)+\frac{\partial g(x)}
{\partial x^\mu}dx^{\mu}\right)~(I+i\lambda A_{\mu}dx^{\mu})~g^{-1}(x)$$
\begin{eqnarray}
~~~~~=g(x+dx)(I+i\lambda A_{\mu}dx^{\mu})g^{-1}(x).
\end{eqnarray}
Obviously there is the equivalence of gauge transformation GT2b (Eq. (5))
and two-point spinor
property of infinitesimal parallel transporter. 

Two-point spinor property for an infinitesimal parallel transporter is crucial 
for the invariance of parallel
transport equation of $\psi$ (Eq. (12)). In terms of infinitesimal
parallel transporter the covariant change of $\psi (x)$, which means the
difference of the field at $x$ and that parallelly transported from $x+dx$,
can be expressed as
\begin{eqnarray}
\delta^{cov}\psi(x)=(I-i\lambda A_{\mu}(x)dx^{\mu})\psi(x+dx)-\psi(x)
=D_{\mu}\psi (x)dx^{\mu}
\end{eqnarray}
Under local gauge transformation it is transformed according to Eq. (3)
and Eq. (19) to
\begin{eqnarray}
\delta^{cov}\psi'(x)=g(x)\left(\frac{\partial}{\partial x^{\mu}}- i\lambda
A_{\mu}(x)\right)
\psi (x)dx^{\mu}.
\end{eqnarray}
Thus the differential equation for the parallel transport of
$\psi(x)$, which can also be express as
\begin{eqnarray}
\lim_{\delta t \rightarrow 0}\delta^{cov}\psi/\delta t=0, 
\end{eqnarray}
is transformed covariantly under gauge transformation, as long as
infinitesimal parallel transporter is a two-point
spinor. This is consistent
with the invariance of the `horizontal
direction' in bundle space under the action of structure group.

\section{Properties of parallel transporter in non-Abelian gauge theory}

From Eq. (17) there are obviously the following three properties of parallel
transporter 
$$  \Phi_{\gamma_2 \circ \gamma_1}( A_{\mu})=\Phi_{\gamma_2} (A_{\mu})
\Phi_{\gamma_1}(A_{\mu}); $$
$$ \Phi_{\bar \gamma}(A_{\mu})=(\Phi_{\gamma}(A_{\mu}))^{-1};$$
$$\Phi_{\gamma}(CA_{\mu}C^{-1})=C\Phi_{\gamma}(A_{\mu})C^{-1}. $$
Here $\bar \gamma$ denotes the inverse path of $\gamma$ and $C$ is a global gauge
transformation.

In this section we will primarily study the  
property of non-Abelian parallel transporter under
local gauge transformation. It was always taken for granted that finite parallel
transporter should be a two-point spinor too, since it can be pieced 
together with infinite number of infinitesimal parallel transporter, which
have been proved to be two-point spinors. Along with the analysis on the
causes
for the false statement, we will give a detailed study
on parallel transport equation, finite and infinitesimal parallel transporter
and their connection under gauge transformations.

\subsection{Investigation into differential equation Eq. (8) and Eq. (12)}

The differential equation that determines the parallel transport
of matter field $\psi(x)
$ (Eqs. (8) and (12)) is of the type
\begin{eqnarray}
\frac{dx}{dt}(t)=W(t)x(t).
\end{eqnarray}
Let us study the behavior of the equation under the transformation
\begin{eqnarray}
x(t)=L(t)y(t).
\end{eqnarray}
Substituting Eq. (24) into Eq. (23), we have
\begin{eqnarray}
\frac{d}{dt}\left(L(t)y(t)\right)=\frac{dL}{dt}(t)y(t)+L(t)\frac{dy}{dt}(t)
=W(t)L(t)y(t).
\end{eqnarray}
Then Eq. (23) is transformed to
\begin{eqnarray}
\frac{dy}{dt}(t)=W'(t)y(t),
\end{eqnarray}
with
\begin{eqnarray}
W'(t)=-L^{-1}(t)\frac{d}{dt}L(t)+L^{-1}(t)W(t)L(t).
\end{eqnarray}
The relation between $W$ and $W'$ is just that of gauge transformation GT2b if
$L^{-1}$ is identified with gauge group element $g$ and, therefore, it is
concluded that
\begin{eqnarray}
\psi'(t)=Pexp\left(\int_{t_0}^{t}dsA'(s)\right)~\psi'(t_0)=
g(t)Pexp\left(\int_{t_0}^{t}dsA(s)\right)~g^{-1}(t_0)\psi'(t_0).
\end{eqnarray}

Seemingly the procedure will lead to the conclusion Eq. (14), but definitely
it has only proved the two-point spinor property of the parallel transporter
when
the field $\psi(t)$ is parallelly transported over a one-dimensional range
$[t_0,t]$, a trivial field theory, because all the operations in the
procedure are stuck to the one-dimensional range $[t_0,t]$ and the integral
in Eq. (28) should be interpreted as a one-variable integral.
In any one-dimensional situation we can always find a local gauge
transformation GT2b that makes gauge field $A(t)$ vanish identically because
the differential equation
\begin{eqnarray}
\frac{dg}{dt}(t)g^{-1}(t)+g(t)A(t)g^{-1}(t)=0
\end{eqnarray}
has definite solutions, then the two-point spinor property holds absolutely 
(see Appendix A).
In fact, a parallel transport path should be regarded as the map $[t_0,t]\to
M$ rather than $[t_0,t]$ itself. The point will be clearly seen through
the discussion in the following subsections. 

A special case involving the solution of Eq. (8) that needs to be
clarified is a periodic $A(s)$.
Without the loss of generality we suppose $A(s+2\pi)=A(s)$.
Then for a parallel transport path,
$[0,4\pi]
\to M$,
we have
$$\Phi(4\pi,2\pi)\Phi(2\pi,0)=Pexp\left(\int^{4\pi}_{2\pi}dsA(s)
\right)~Pexp\left(\int^{2\pi}_{0}dsA(s)\right).$$ 
Its term of the order $(i\lambda)^2$ is
$$\int^{2\pi}_{0}dsA(s)\int^{s}_{0}ds'A(s')+
\int^{4\pi}_{2\pi}dsA(s)\int^{s}_{2\pi}ds'A(s')+
\int^{4\pi}_{2\pi}dsA(s)\int^{2\pi}_{0}dsA(s)$$
\begin{eqnarray}
=\int^{4\pi}_{0}dsA(s)\int^{s}_{0}ds'A(s')+
(\int^{2\pi}_{0}dsA(s))^2,
\end{eqnarray}
with the presence of the periodic function $A(s)$. 
If the $(i\lambda)^2$ term of $\Phi(4\pi,0)=Pexp\int^{4\pi}_0dsA(s)$
should also
be formally given as
$\int^{4\pi}_{0}dsA(s)\int^{s}_{0}ds'A(s')$ according to Eq.
(10), the integral factor $\int^s_0ds'A(s')$ in it is indefinite due to
the period $2\pi$ of $A(s)$.
To guarantee a definite group composition law:
$$\Phi(4\pi,2\pi)\Phi(2\pi,0)=\Phi(4\pi,0),$$ 
we must specify that the parallel transporter constructed by
a periodic $A(s)$ should be expressed as the product of those along the 
monotonic one, i.e. $\Phi(2\pi n,0)=\Phi^n(2\pi,0)$
in the example.

\subsection{Investigation into infinite product Eq. (17)}

To see if the two-point spinor property of finite parallel transporter, a
non-local operator, can be
obtained through piecing together two-point spinor property of local
infinitesimal
parallel transporter, we need to study the infinite product form of finite
parallel transporter Eq. (17), since we have proved that an infinitesimal 
parallel transporter
will be a two-point spinor only when it is in the form of one-order expansion.
The transport path, $[t_0,t]\to M$, is 
divided into countably
infinite small range $[t_i,t_{i+1}]$, for $i=0,1,\cdots,n$, then a finite 
parallel
transporter after gauge transformation GT2 becomes the following path-ordered infinite 
product:
\begin{eqnarray}
\Phi(A'_{\mu}(x); t,t_0)=(I+A'(t_{n-1})\Delta t)\cdots
(I+A'(t_1)\Delta t)(I+A'(t_0)\Delta t),
\end{eqnarray}
where $A'(t_i)=i\lambda A'_{\mu}\left(x(t_i)\right)dx^{\mu}/dt(t_i)$, and
$\Delta t\rightarrow 0$.
With the two-point spinor property of an infinitesimal parallel transporter in the
form of the one-order expansion along the transport path,
it is equal to
$$\left(U(t_{n-1})+\Delta U(t_{n-1})\right)
(I+A(t_{n-1})\Delta t)U^{-1}(t_{n-1})\left
(U(t_{n-2})+\Delta U(t_{n-2})\right)
(I+A(t_{n-2})\Delta t)\cdots$$
$$\cdots \left(U(t_0)+\Delta U(t_0)\right)(I+A(t_{0})\Delta t)U^{-1}(t_0),$$
where $U(t)=g(x(t))$.
If we take
\begin{eqnarray}
U(t_{i+1})=U(t_{i})+\Delta U(t_{i})=U(t_{i})+\frac {dU}{dt}(t_i)\Delta t
\end{eqnarray}
for $i=1,2,\cdots ,n-1$, then all the group elements in the middle will be 
canceled in couples and a 
two-point spinor $\Phi(A_{\mu}(x);t,t_0)$ will be obtained.

To check the validity of the argument, we can study the relation of the gauge
group elements at the two end points of the transport path.
The simplest case is a smooth group element function $U(t)$ on the
transport path,
i.e. there are infinite-th partial derivatives with respect to
spacetime variables at
each point of
the path, $[t_0,t]\to M$.
According to Eq. (32), the gauge group elements are related by 
iterative one-order expansion along the transport path, so we obtain the
following relation (see Appendix B) of the gauge group elements at the end
points
of a smooth transport path:
$$ U(t)= U(t_{n-1})+\frac {dU}{dt}(t_{n-1})\Delta t $$
$$= U(t_{n-2})+2\frac {dU}{dt}(t_{n-2})\Delta t+\frac
{dU^2}{dt^2}(t_{n-2})(\Delta t)^2= \cdots$$
\begin{eqnarray}
=\sum_{k=0}^{n}\left(
\begin{array}{c}
n \\
k
\end{array}
\right)(\frac{d}{dt})^kU(t_0)(\Delta
t)^k,
\end{eqnarray}
where $\Delta t=(t-t_0)/n$.
When $n\rightarrow \infty$, it reduces to the Taylor expansion in variable $t$:
\begin{eqnarray}
U(t)=U(t_0)+(t-t_0)U^{\prime}(t_0)+\frac{1}{2!}(t-t_0)^2U^{(2)}(t_0)+\cdots.
\end{eqnarray}
Changing the variable of differentials $t$ to spacetime variables $x$,
we have 
\begin{eqnarray}
g(x)=g(x_0)+\sum\limits_{k=1}^{\infty}\frac{1}{k!}\left((x^{\mu}-x_0^{\mu})
\frac{\partial}{\partial x^{\mu
}}\right)^kg(x_0).   
\end{eqnarray} 
This is not definitely consistent with the Taylor expansion of $g(x)$ directly with
respect to
spacetime variables $x$, if $g(x)$ is only a $C^{k}(k<\infty)$ function,
a general local gauge transformation, on the subset of $M$, into which the
parallel transport path is embedded. The relation of the Taylor expansion
in path parameter
$t$ and spacetime variables $x$ is given in Appendix C.
From the above discussion it can be seen that it is improper to treat the
parallel transport
of matter field on $M$ as a one-dimensional problem in the parameter space
$[t_0,t]$.

\subsection{Condition for the preservation of parallel transport under
gauge transformation}

First we start with the following equation in the parameter space
$[t_0,t]$:
$$\psi(t)-\psi(t_0)=Pexp\left(\int^t_{t_0}dsA(s)\right)\psi(t_0)-\psi(t_0)$$
$$=\int^{t}_{t_0}dsA(s)\left(I+\int^s_{t_0}ds'A(s')+\cdots\right)\psi(t_0)$$
\begin{eqnarray}
=\int^{t}_{t_0}dsA(s)Pexp\left(\int^s_{t_0}ds'A(s')\right)\psi(t_0)
=\int^{t}_{t_0}dsA(s)\psi(s),
\end{eqnarray}
where $A(s)=i\lambda A_{\mu}(x(s))dx^{\mu}(s)/ds$. 
It is the integral equation of the parallel transport of fermions and is
equivalent to Eq. (12). If the concerned parallel transporter is
a two-point spinor, this equation should transform covariantly 
under
gauge transformation. On the spacetime manifold M, the transformed Eq.
(36) after a coordinate transformation in principal bundle
space (Eq. (2)) is
$$\psi'(x)=\psi'(x_0)+i\lambda\int^x_{x_0}dz^{\mu}A'_{\mu}(z)\psi'(z)$$
\begin{eqnarray}
=g(x_0)\psi(x_0)+\int^x_{x_0}dz^{\mu}(i\lambda
g(z)A_{\mu}(z)+\partial_{\mu}g(z))\psi(z),
\end{eqnarray}
if $\psi(z)$ produced by parallel transport, a $C^{\infty}$ extension of
$\psi(s)$ to M (the
transport curves under consideration are embedding ones), transforms
covariantly.  
To treat the integral in the equation as a one-parameter integral in
$[t_0,t]$, we have 
$$\psi'(t)=U(t_0)\psi(t_0)+\int^t_{t_0}ds\left(U(s)A(s)\psi(s)+
\frac{d}{ds}(U(s)\psi(s))\right)-\int^t_{t_0}ds~U(s)\frac{d}{ds}\psi(s)$$
\begin{eqnarray}
=U(t)\psi(t)=U(t)(\psi(t_0)+\int^{t}_{t_0}dsA(s)\psi(s)),
\end{eqnarray}
where $U(t)=g(x(t))$ and Eq. (12) has been considered.
So the integral equation for the parallel
transport of fermions appears to transform covariantly under gauge
transformation.

However, as we will clarify as follows, such a treatment actually leads to
the contradiction with facts. Let's study a situation described by the
figure 1.
\begin{figure}
\label{q40}
\epsfysize=3in
\hspace{4cm}
\epsffile{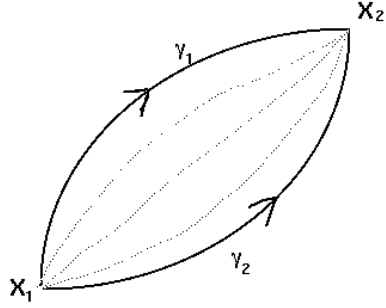}
\caption[]{A spinor $\psi_0$ at $x_1$ is parallelly transported to $x_2$
along a family of curves connecting the two points.}
\end{figure}
The corresponding integral equations on $\gamma_1$ and $\gamma_2$ are
given as follows:
\begin{eqnarray}
Pexp\left(\int_{\gamma_1}dsA(s)\right)\psi(t_0)-\psi(t_0)=
\int_{\gamma_1}dsA(s)\psi(s),
\end{eqnarray}
\begin{eqnarray}
Pexp\left(\int_{\gamma_2}dsA(s)\right)\psi(t_0)-\psi(t_0)=\int_{\gamma_2}dsA(s)\psi(s).
\end{eqnarray}
We have the following relation after subtracting Eq. (40) from Eq. (39)
$$Pexp\left(\int_{\gamma_1}dsA(s)\right)\psi(t_0)-Pexp\left
(\int_{\gamma_2}dsA(s)\right)\psi(t_0)
=\int_{\gamma_1}dsA(s)\psi(s)-\int_{\gamma_2}dsA(s)\psi(s)$$
\begin{eqnarray}
=\int_{\bar\gamma_2\circ\gamma_1}dsA(s)\psi(s)
\end{eqnarray}
The left hand side of it can be rewritten as
$$Pexp\left(\int_{\gamma_1}dsA(s)\right)~(I-Pexp\left(\int_{\bar\gamma_1}dsA(s)
\right)
\times Pexp\left(\int_{\gamma_2}dsA(s)\right))\psi(t_0)$$
\begin{eqnarray}
=Pexp\left(\int_{\gamma_1}dsA(s)\right)(-\int_{\bar\gamma_1\circ\gamma_2}
dsA(s)\psi(s)), 
\end{eqnarray}
if Eq. (36) is considered.
The final equation is therefore given as
\begin{eqnarray}
(Pexp\left(\int_{\gamma_1}dsA(s)\right)-I)
\int_{\bar\gamma_2\circ\gamma_1}dsA(s)\psi(s)=0.
\end{eqnarray}
It indicates a trivial result for the parallel transport of $\psi_0$
in the parameter space $[t_0,t]$, which goes against facts. 
Thus we have shown again that the
differential
form $A(t)dt$ in the integrals of all the concerned equation have to be
interpreted
not only as the one-form in the parameter space $[t_0,t]$ but also as the
one-form
$i\lambda A_{\mu}(x)dx^{\mu}$ on spacetime manifold M itself.

Next we will reveal the actual condition for the two-point spinor property
of parallel transporter. We consider the situation in the above figure
again and set up a coordinate of a homotopic curve family, $x^{\mu}(t,s)$,
on the surface surrounded by $\gamma_1=x^{\mu}(t,0)$ and
$\gamma_2=x^{\mu}(t,1)$. At the two end points there is
$\frac{\partial}{\partial s}x^{\mu}(0, s)=\frac{\partial}{\partial 
s}x^{\mu}(1, s)=0$. Due to the
parallel transport of $\psi_0$ along
the family of curves, there is a fermion field $\psi(t,s)$ on the surface.
If the parallel transporters $\Phi_{\gamma}$ are two-point spinors,
the difference between the transported fermions along $\gamma_1$ and
$\gamma_2$ respectively to $x_2$ will transform
covariantly under gauge transformation:
\begin{equation}
\Delta\psi'(x_2)=g(x_2)\Delta\psi(x_2)=
U(1,s)(\Phi_{\gamma_2}-\Phi_{\gamma_1})\psi_0
\end{equation}
According to the Stokes theorem in the surface parameter space $\{t,s\}$,
$$\Delta\psi(x_2)=
ig\int^1_0A_{\mu}(x(t,1))\psi(t,1)\frac{\partial x^{\mu}}
{\partial
t}dt-ig\int^1_0A_{\mu}(x(t,0))\psi(t,1)\frac{\partial x^{\mu}}{\partial
t}dt$$
$$=ig\oint_{\bar\gamma_1\circ\gamma_2}A_{\mu}(x(t,s))\psi(t,s)\frac{\partial
x^{\mu}}{\partial t}dt$$
\begin{eqnarray}
=\int^1_0ds\int^1_0dt~ig\{ 
\frac{\partial}{\partial t}(A_{\mu}(x(t,s))\psi(t,s)\frac{\partial x^{\mu}}
{\partial s})
-\frac{\partial}{\partial s}
(A_{\mu}(x(t,s))\psi(t,s)\frac{\partial x^{\mu}}{\partial t})\}.
\end{eqnarray}
In the transformed coordinate in principal bundle space (after the action
in Eq. (2)), $\Delta\psi'(x_2)$ is expressed similarly
as
$$\Delta\Psi'(x_2)=ig\oint_{\bar\gamma_1\circ\gamma_2}A'_{\mu}(x(t,s))\psi'
(t,s)\frac{\partial x^{\mu}}{\partial t}dt$$
$$=\int^1_0ds\int^1_0dt~ig\{
\frac{\partial}{\partial t}
(g(x(t,s))A_{\mu}(x(t,s))\psi(t,s)\frac{\partial x^{\mu}}{\partial s})$$
$$-\frac{\partial}{\partial s}
(g(x(t,s))A_{\mu}(x(t,s))\psi(t,s)\frac{\partial x^{\mu}}{\partial t})\}$$
\begin{eqnarray}
+\int^1_0ds\int^1_0dt\{\frac{\partial}{\partial t}(\partial_{\mu}g(x(t,s))
\psi(t,s)\frac{\partial x^{\mu}}{\partial s})-
\frac{\partial}{\partial s}(\partial_{\mu}g(x(t,s))
\psi(t,s)\frac{\partial x^{\mu}}{\partial t})\},
\end{eqnarray}
if $\psi(t,s)$ is supposed to transform covariantly.
Only when the $\Delta\psi'(x_2)$ calculated in the transformed bundle
coordinate (Eq. (46)) and that transformed from $\Delta\psi(x_2)$ in 
Eq.(45) by
$g(x_2)$ are equal, will the parallel transporters under consideration be
truly two-point spinors. In other words it is the necessary condition for
the 
two-point spinor property of parallel transporter. From
explicit calculation, however, their
equivalence
is true only in the following two situations:

1) $g(x)=const$, a global gauge transformation GT1;

2) a pure gauge situation with $F_{\mu\nu}(x)=0$ identically. 

In the second case,
the transported fermion satisfies the following partial differential
equation.
\begin{eqnarray}
\partial_{\mu}\psi(x)=igA_{\mu}(x)\psi(x),
\end{eqnarray}
then the general covariance of the integral equation for the
parallel transport of fermions can be restored by means of integral by
parts with respect to $\partial_{\mu}$ in Eq. (37).  It is in contrast to
the general covariance of the differential equation (Eqs. (12) and (22))
for the parallel
transport of fermions that the general covariance of its equivalent
integral
formalism (Eq. (36)) under gauge transformation is conditional.
As has been given by a similar result of the parallel transport of vector
field in tangent bundle space based on a Riemannian spacetime [13, p167],
the condition $F_{\mu\nu}(x)=0$ identically is the sufficient and
necessary
condition for 
the equivalence of the ordinary
differential equation Eq. (12) and the partial differential equation Eq.
(47). i.e. that for the integrability of parallel
transporter $\Phi_{\gamma}$, and, through the previous discussion, it
is also the sufficient and necessary condition for the preservation of
parallel transport of fermions under a general $(C^1)$ local gauge
transformation.

\subsection{Direct verification}

The restriction on preservation of parallel transport of fermions in a
non-trivial gauge field can be
shown directly, if we treat the integrals in the following supposed
equation as line integral in spacetime manifold M rather than the
one-parameter integral in the parameter space $[t_0,t]$ itself.
It is done by comparing
the $(i\lambda)^n$ order terms on the both sides of the supposed equation:
\begin{eqnarray}
Pexp\left(i\lambda\int_{x_0}^{x}dz^{\mu}A'_{\mu}(z)\right)=
g(x)Pexp\left(i\lambda\int_{x_0}^{x}
dz^{\mu}A_{\mu}(z)\right)~g^{-1}(x_0),
\end{eqnarray}
where
$$g(x)=exp~(i\lambda\sum \limits^{dimG}_{a=1}\omega^{a}(x)T^a)~=
exp~(i\lambda M(x))
=I+i\lambda M(x)+\cdots,$$
and
$$g^{-1}(x_0)=exp~(-i\lambda\sum\limits^{dimG}_{a=1}\omega^{a}(x_0)T^a)
=exp~(-i\lambda M(x_0)).$$

After the gauge transformation of $A_{\mu}(z)$ is substituted
into the left hand side of Eq. (48), we compare the terms of the order
$(i\lambda)^2$.
The terms
of the order with the permutation MA\footnote[3]
{M means the part containing factors such as $dM(z)$ and $M(z)$, and A means
those with $A(z)$ and $\int^z_{x_0}dz^{\mu}A_{\mu}(z)$.} 
on the left hand side of Eq. (48) are given as follows:
$$(i\lambda)^2\int^x_{x_0}dz^{\mu}M(z)A_{\mu}(z)+(i\lambda)^2\int^{x}_{x_0}
dM(z)\int^z_{x_0}(dz^{\mu})'A_{\mu}(z')=
(i\lambda)^2\int^x_{x_0}dz^{\mu}M(z)A_{\mu}(z)$$
\begin{eqnarray}
-(i\lambda)^2\int^{x}_{x_0}M(z)
d\left(\int^z_{x_0}(dz^{\mu})'A_{\mu}(z')\right)~+(i\lambda)^2M(x)\int^x_{x_0}dz^{\mu}
A_{\mu}(z).
\end{eqnarray}
If the integrals here are interpreted 
as one-variable integrals in $[t_0,t]$, i.e.
$i\lambda\int^{z}_{x_0}(dz^{\mu})'A_{\mu}(z')=\int^s_{t_0}ds'A(s')$
and
$A(s)ds=i\lambda A_{\mu}(z)dz^{\mu}$,
there
is \begin{eqnarray}
d\left(\int^s_{t_0}ds'A(s')\right)
= A(s)ds
\end{eqnarray}
as an identity, and the two sides of Eq. (48) will agree up to all orders, 
as it is
always true for the one-dimensional situation. 
However, as a matter of fact, Eq. (50) implies
\begin{eqnarray}
d\left(i\lambda\int^z_{x_0}(dz^{\mu})'A_{\mu}(z')\right)= 
i\lambda A_{\mu}(z)dz^{\mu},
\end{eqnarray}
which is true on $M$ only when $A_{\mu}(z)dz^{\mu}$ is an exact form, i.e.
$\partial_{\mu}A_{\nu}(z)
-\partial_{\nu}A_{\mu}(z)=0$ identically (see e.g. Ref.[14, p.~10]). 
Therefore, the consistency for the validity of Eq. (48) 
in $[t_0,t]$ with that on $M$ 
doesn't always hold, if the gauge field under consideration is a
non-trivial one.

\section{Abelian gauge theory and Aharonov-Bhom effect}

 The gauge field theory proper with gauge group $U(1)$ is a special type in
our discussion. The quantized $U(1)$ gauge field (Quantum Electrodynamics)
well describes electromagnetic interaction in nature. The variation of its
Lagrangian density 
\begin{eqnarray}
L(x)=-\frac{1}{4} F_{\mu\nu}(x) F^{\mu\nu}(x)+\bar \psi(x)
\left( i \gamma^\mu D_\mu - m \right) \psi(x) 
\end{eqnarray}
with respect to field $\bar\psi(x)$ leads to Dirac equation
\begin{eqnarray}
\left( i\gamma^\mu D_\mu - m \right) \psi(x)=0, 
\end{eqnarray}
where $D_{\mu}=\partial_{\mu}-ieA_{\mu}$. In the region of spacetime where field 
strength $F_{\mu\nu}(x)$ identically vanishes, it can be reduced to the 
equation of free field 
\begin{eqnarray}
\left( i \gamma^\mu \partial_\mu - m \right) \psi_f(x)=0 
\end{eqnarray}
through a phase factor called `Schwinger String' [15]:
\begin{eqnarray}
\psi(x)=exp\left(ie\int_{x_0}^{x} A_{\mu}(z)dz^{\mu}\right)~\psi_f(x).
\end{eqnarray}

In the non-relativistic limit, this mathematical transformation corresponds to
a physical phenomena. It is the famous Aharonov-Bohm
effect that
directly demonstrates the effect of gauge potential $A_{\mu}$ on electron
field [16,17]. The experiment is arranged to allow two beams of electron
from a
single source to pass either side of a round coil and impinge on a screen
behind. The magnetic field is restricted to within the coil so
that field tensor $F_{\mu \nu}$ in where electrons pass is identically zero.
If the origin of the space coordinate is chosen at the center of the coil,
the one-form of the gauge field will be $A_{\mu}(x)dx^{\mu}=-
y/(x^2+y^2)dx+x/(x^2+y^2)dy$, which cannot be transformed to zero
by a $C^{\infty}$ gauge transformation because it is not the differential
of a function over $M/(0,0)$ [14, p.~6].
Hence we have a double-connected region on $M$.
The interference pattern at each point $x$ on the screen is determined by
the
amplitude,
$$\vert \psi_{f_1}(x)exp\left(ie\int_{\gamma_{1}}dz^{\mu}A_{\mu}(z)\right)+
\psi_{f_2}(x)exp\left(ie\int_{\gamma_{2}}dz^{\mu}A_{\mu}(z)\right)\vert^2$$
\begin{eqnarray}
=\vert\psi_{f_1}(x)exp\left(ie/2\int\int_{\Sigma}d\sigma^{\mu\nu}(z)
(\partial_{\mu}A_{\nu}(z)-\partial_{\nu}A_{\mu}(z)\right)~+\psi_{f_2}(x)\vert^2,
\end{eqnarray}
where the integral domain is over the cross section of the coil, $\Sigma$,
since the field strength vanishes identically outside. The amplitude is
invariant under
the local gauge transformations: 
\begin{eqnarray}
\psi'(x)=exp\left(ie\alpha(x)\right)\psi(x),
\end{eqnarray}
\begin{eqnarray}
A'_{\mu}(x)=A_{\mu}(x)+\partial_{\mu}\alpha(x),
\end{eqnarray}
because the field strength
$F_{\mu\nu}(x)$
is gauge invariant.

If a double-connected region can be realized
in a non-Abelian situation, i.e. the magnetic field in the experiment would be
taken place by some non-Abelian gauge field and the electron field by
some fermion field coupling to the non-Abelian gauge field, the
corresponding amplitude of Eq. (56) will be
$$\vert Pexp\left(i\lambda\int_{\gamma_{1}}dz^{\mu}A_{\mu}(z)\right)\psi_{f_1}(x)+
Pexp\left(i\lambda\int_{\gamma_{2}}dz^{\mu}A_{\mu}(z)\right)\psi_{f_2}(x)
\vert^2$$
\begin{eqnarray}
=\vert\psi_{f_1}(x)\vert^2+\vert\psi_{f_2}(x)\vert^2+\vert\psi^{\dag}_{f_2}(x)
Pexp\left(i\lambda
\oint_{\Gamma} dz^{\mu}A_{\mu}(z)\right)\psi_{f_1}(x)+h.c.\vert^2,
\end{eqnarray}
where $\Gamma=\gamma_1\cup\bar\gamma_2$ with the point $x_0$ as the
initial and final point. The phase factor here can be
calculated with the help of `non-Abelian
Stokes
theorem' [18,19]: 
$$Pexp\left(i\lambda\oint_{\Gamma}
dz^{\mu}A_{\mu}(z)\right)~=Pexp\left(i\lambda\int_y^{x_{0}}
A_{\mu}(z)dz^{\mu}\right)
Pexp\left(i\lambda/2\int\int_{\Sigma} 
d\sigma^{\mu\nu}(z)F_{\mu\nu}(y,z)\right) $$
\begin{eqnarray}
\times Pexp\left(i\lambda\int_{x_0}^yA_{\mu}(z)dz^{\mu}\right),
\end{eqnarray}
where $y$ is an arbitrary
reference point on
$\Sigma$, and
$$F_{\mu\nu}(y,z)=Pexp~(i\lambda\int_z^y A_{\mu}dx^{\mu})F_{\mu\nu}(z)
Pexp~(i\lambda\int_y^z A_{\mu}dx^{\mu}).$$ 
With regard to Eq. (59), the contribution to
the amplitude involving the phase factor around the coil becomes
$$\vert\psi^{\dag}_{f_2}(x)g^{-1}(x)Pexp\left(i\lambda\int_{\Gamma'}A'_{\mu}(z)dz^{\mu}
\right)g(x)\psi_{f_1}(x)+h.c.\vert^2,$$
after
the local gauge transformations GT2a and GT2b induced by
$T_{\alpha\beta}(x)$.
Here $\Gamma'=C\cup S^1\cup \bar C$, with $S^1$ the boundary of the coil
and $C$ arbitrary path connecting $x_0$ and $S^1$.  
It is not equal to the corresponding part in Eq. (59) because
$x\not\in \Gamma'$ and the local gauge transformations
of
the phase factor only involve the gauge group elements on the path
$\Gamma'$, so the interference pattern is therefore invariant
only under global gauge
transformation GT1.
Without the consideration of the confinement of fermions in
non-Abelian gauge field theories, this imaginary experiment demonstrates
that only Abelian gauge field theory is a
perfect locally invariant gauge theory in the sense of the definition in
Sect.(2), i.e. a
coordinate-free
description of physics can be realized only in $U(1)$ bundle space.

\section{Discussions}
Both in model construction [20-22] and
lattice simulation [23], non-perturbative approaches to QCD widely
involves the application of parallel
transporter. The closed parallel transporter (Wilson loop)
is an important tool for the study of the non-perturbative phenomena such as
the confinement of quarks (see, e.g. Ref.[9, chap.~5]). However, the
explicit 
broken of the local gauge
symmetry of the quantities constructed with this non-local operator,
parallel transporter, was not understood before. In the perturbative domain,
an infinitesimal parallel transporter 
can be regarded as a good approximation to
the real one, because of the aymptotic freedom property for 
non-Abelian gauge theories. Its two-point spinor property preserves the local
gauge invariance GT2 in the perturbative domain. Whether the loss of two-point
spinor property of parallel transporter in non-perturbative domain implies
some physics requires our further study.

Another issue closely related to gauge symmetry is general covariance in
the
geometrized theories of gravitation, e.g. general relativity. In fact the
concept of
gauge invariance originated from H. Weyl's attempt [24] to unify
gravitation and
electromagnetism in the framework of Riemannian geometry. General relativity
was first treated as a gauge theory of orthonormal frame bundle with the
homogeneous Lorentz group 
as the structure group in Ref.[25]. For a conceptual
development of gauge concept and geometrization of fundamental interactions,
see Ref.[26]. The largest symmetry in spacetime theories is the invariance
of the
local field equations under local coordinate transformation, and people
used to
believe that this group of differmorphism comprises the gauge freedom of
any theory formulated in terms of tensor fields on a spacetime manifold $M$,
and
therefore all differmorphic models of any spacetime theory represent one and
the same physical situation. Hence the term {\bf general covariance} for a
physical theory usually refers to the invariance of local field equations
under coordinate 
transformation in tangent bundle space.
For a genuine equivalence of physics, however, 
a tensor field produced by parallel transport should also be covariant 
under
the related transformation law. 
For example, if a physical process involves the parallel transport of a vector
field $n^{\mu}(x)$
between two points, say $x_1$ and $x_2$, on spacetime manifold,
the parallel transporter, $Pexp\left(-\int_{x_1}^{x_2} \Gamma_{\mu
\nu}^{\sigma}
(x)
dx^{\nu}\right)$, of the
vector field must be a two-point tensor [5,6]. From the
argument in the previous discussion it is true only when
the curvature tensor of the spacetime manifold vanishes
identically. Generally
speaking, the parallel transport of
tensor fields should be an `absolute element' [27] in any spacetime
theory, i.e. the concerned spacetime transformation should map the
parallel transporter as a geometrical object
in one model to the corresponding ones 
in all its equivalent models, so the models should be geometrically
rather than topologically equivalent. 
This requirement imposes stronger condition for the general
covariance of
a theory than that for the form invariance of local Lagrangian density. 
Following the argument in this paper we can show
the loss of the property in the general situation through the analysis of 
parallel transporters
for tensor fields. When
studied from integral point of view, the two most fundamental symmetries,
the general covariance in principal and tangent bundle space, are shown to
be much more
restricted than what we see from differential point of view.

\vskip 10mm
\noindent
{\bf\large Acknowledgments}. The work is partially supported by National
Nature
Science Foundation of China under Grant 19677205.

\newpage
\noindent
{\bf\large Appendix~A}
\renewcommand{\theequation}{A.\arabic{equation}}
\setcounter{equation}{0}

In this appendix we prove the two-point spinor property for parallel
transporter when the gauge field $A_{\mu}(x)$ can be transformed to zero
identically. It is in fact to prove
\begin{eqnarray}
Pexp\left(i\lambda\int^x_{x_0}dz^{\mu}A_{\mu}(z)\right)~
=Pexp\left(\int^x_{x_0}dz^{\mu}\partial_{\mu}g(z)g^{-1}(z)\right)
=g(x)g^{-1}(x_0),
\end{eqnarray}
where $g(z)=exp(i\lambda~\omega^a(z)T^a)= expi\lambda M(z)$.
We have
\begin{eqnarray}
\partial_{\mu}g(z)g^{-1}(z)=\partial_{\mu}\left(\sum\limits_{k=0}^{\infty}\frac{1}
{k!}(i\lambda
M(z))^k\right)~\left(\sum\limits_{k=0}^{\infty}\frac{(-1)^k}{k!}(i\lambda
M(z))^k\right)~~.
\end{eqnarray}
Substituting it into (A.1). we obtain infinitely many terms containing
$M(z)$ as the integral variable, such as $\int
d(M^l/l!)((-1)^n/n!)M^n)$.
All these terms can be grouped into the (n+1)-element sets of the order
$(i\lambda)^{l+n+x}$ in the form:
$$\int\cdots\int^{s_{i-1}}_{t_0}d(M^l/l!)\frac{(-1)^k}{k!}M^k
\int^{s_{i}}_{t_0}d\left(\frac{1}{(n-k)!}M^{n-k}\right)~\int\cdots,$$
for $k=0,1,\cdots,n$, where the common factors $d\left(M^l/l!\right)$
contributes the order $(i\lambda)^l$, and the common part $\int\cdots$ the order
$(i\lambda)^x$.
We find that the terms in such a group cancel all together, because
$\sum\limits_{k=0}^{n}(-1)^{k}(1/k!)(1/(n-k)!)=0$
from the identity
\begin{eqnarray}
(1-1)^n=n!\left(\sum\limits_{k=0}^{n}(-1)^{k}\frac{1}{k!}
\frac{1}{(n-k)!}\right)=0.
\end{eqnarray}
In this way only the terms on the right hand side of B.(1) will be left.

\vspace {4mm}
\noindent
{\bf\large Appendix~B}
\renewcommand{\theequation}{C.\arabic{equation}}
\setcounter{equation}{0}

Here Eq. (33) is proved by the induction on $n$:
If $k=1$, then for any differentiable function $g(t)$ there is 
\begin{eqnarray}
g(t+\Delta t)=g(t)+dg/dt(t)\Delta t
\end{eqnarray}
under one-order approximation.
Suppose Eq. (33) holds for $k=n-1$. Then we have
\begin{eqnarray}
g\left(t+(n-1)\Delta t\right)~=\sum \limits_{k=0}^{n-1}\left(
\begin{array}{c}
n \\
k
\end{array}
\right)~(d/dt)^kg(t_0)(\Delta t)^k,
\end{eqnarray}
and
\begin{eqnarray}
dg/dt\left(t+(n-1)\Delta t\right)~=\sum \limits_{k=0}^{n-1}\left(
\begin{array}{c}
n \\
k
\end{array}
\right)~(d/dt)^kdg/dt(t_0)(\Delta t)^k.
\end{eqnarray}
When $k=n$, it immediately follows that
\begin{eqnarray}
g\left(t+n\Delta t\right)~=g\left(t+(n-1)\Delta
t\right)~+dg/dt\left(t+(n-1)\Delta t\right)\Delta t.
\end{eqnarray}
Substituting B.2 and B.3 into B.4 and considering the relation
\begin{eqnarray}
\left(
\begin{array}{c}
n \\
k
\end{array}
\right)
=\left(
\begin{array}{c}
n-1 \\
k-1
\end{array}
\right)
+
\left(
\begin{array}{c}
n-1 \\
k
\end{array}
\right),
\end{eqnarray}
we obtain Eq. (33) for $k=n$ after the rearrangement of the terms.

\vspace{4mm}
\noindent
{\bf\large Appendix~C}
\renewcommand{\theequation}{C.\arabic{equation}}
\setcounter{equation}{0}

First we derive Eq. (35) from Eq. (34).
Remember that we suppose $g(x)$ has infinite-th partial derivatives at
each point on the path $\Gamma$, $[t_0,t]\to M$. After the variable $t$ of
the
differentials is changed to $x$, there
are
$$\frac{dU}{dt}(t)=\frac{dx^{\mu}}{dt}(t)\partial_{\mu}g(x), $$
$$\frac{d^2U}{dt^2}(t)=\frac{d^2x^{\mu}}{dt^2}(t)\partial_{\mu}g(x)+
\frac{dx^{\mu}}{dt}(t)\frac{dx^{\nu}}{dt}(t)\partial_{\mu}\partial_{\nu}g(x),
$$
etc. Substituting these results into Eq. (34), we obtain the coefficient
of
$\partial_{\mu}g(x_0)$ as follows:
$$ (t-t_0)\frac{dx^{\mu}}{dt}(t_0)+\frac{1}{2!}(t-t_0)^2\frac{d^2x^{\mu}}
{dt^2}(t_0)+\cdots=x^{\mu}-x_0^{\mu},$$
since the path is a smooth one. All the terms of the higher derivatives in
Eq. (35) are obtained in the same way.

On the other hand, if there are continuous n-th partial derivatives
of $g(x)$ at each point of
$W=\prod\limits^n_{\mu=1}(x^{\mu}-x^{\mu}_0)\subset M$,
then it can be expanded to the n-th order of $(x^{\mu}-x^{\mu}_0)$ with
respect to the spacetime variables. 
However, with some points $z\not\in\Gamma$ but $\in W$, at which the
continuous
partial derivatives exist only up to the $m(<n-1)$-th,
$g(x)$ cannot be certainly expanded in the form:
$$g(x)=g(x_0)+\sum\limits_{k=1}^{n-1}\frac{1}{k!}\left((x^{\mu}-x_0^{\mu})
\frac{\partial}{\partial
x^{\mu}}\right)^kg(x_0)+\frac{1}{n!}\left((x^{\mu}-x_0^{\mu})
\frac{\partial}{\partial x^{\mu
}}\right)^ng(\xi),$$
where $\xi$ is some point in $W$.

As an example, we perform the local gauge transformation
$g(x,y)=exp~(\sum\limits^{dimG}_{a=1}(\alpha^a(x,y))^{\frac{4}{3}}T^a)$,
where $\alpha^a(x,y)$ are $C^{\infty}$ functions, on two-dimensional
plane X-Y. $g(x,y)$ is a $C^1$ function over the domain $W=[0,1]\times [0,1]$ if
there are $\gamma^a\cap W\neq\emptyset$
for some curves $\gamma^a$ determined by the equations,
$\alpha^a(x,y)=0$.
Consider a smooth path $\Gamma$ connecting the point $(0,0)$ to $(1,1)$,
with $\Gamma\cap\gamma^a=\emptyset$ for all $\gamma^a$s. Then $U(t)$ is
$C^\infty$
on $\Gamma$.
From the Taylor expansion of $U(t)$ we have
$$g(1,1)=g(0,0)+\sum\limits_{k=1}^{\infty}\frac{1}{k!}
\left(\frac{\partial}{\partial x}+\frac{\partial}{\partial y}\right)
^kg(0,0),$$
whereas the Taylor expansion of a $C^1$ function $g(x,y)$ on $W$ only
definitely gives
$$g(1,1)=g(0,0)+(\frac{\partial}{\partial x}+\frac{\partial}
{\partial y})g(\xi_1,\xi_2),$$
where $(\xi_1,\xi_2)$ is some point in $W$.                                   
Therefore, it is concluded that the coincidence of the two Taylor
expansions
Eq. (34) and Eq. (35) requires the same behavior of $g(x)$ on
$\Gamma$ and $W$.

\end{document}